\documentclass{article}
\usepackage{epsfig}

\newcommand{\be}{\begin{equation}}
\newcommand{\ee}{\end{equation}}

\begin{document}
   \title{Dynamics of point Josephson junctions in a microstrip line}
   \author{  J.G. Caputo and L. Loukitch  \\
   {\normalsize \it Laboratoire de Math\'ematiques, INSA de Rouen} \\
   {\normalsize \it B.P. 8, 76131 Mont-Saint-Aignan cedex, France }
   \date{\today}}

\maketitle
\begin{abstract}
We model the dynamics of point Josephson junctions in a 1D microstrip
line using a wave equation with delta distributed sine nonlinearities.
The model is suitable for both low T$_c$ and high T$_c$ systems (0 and
$\pi$ junctions).
For a single junction in the line, we found two limiting behaviors: the
ohmic mode where the junction acts as a pure resistor which stops
waves and separates the cavity and the junction mode where the wave
is homogeneous throughout the strip. This classification allows to bound the
IV curves of the system. Two junctions in a strip give generally ohmic
modes and combined junction/ohmic modes and yield information about the
behavior with an array with many junctions. Finally we use this analysis
to understand the many junction case for 0 and $\pi$ junctions and the
effect of an external magnetic field.
\end{abstract}

\section{Introduction}

Type I Superconductors are characterized by the phase of a complex
number, the macroscopic order parameter. Josephson \cite{josephson} 
showed that when 
two such superconductors are very close, electrons can tunnel from one
to the other according to the relations
\be\label{joseph}
V = {\Phi_0 \over 2 \pi}{d \phi \over dt}~~, ~~ I = s J_c 
\sin(\phi)~,\ee
where $V$ and $I$ are respectively the voltage and current across 
the barrier, $\phi$ is the difference between the phases in 
the two superconductors, $s$ is the contact surface, $J_c$ is the 
critical current density and
$\Phi_0$ is the flux quantum. 
The two Josephson relations together with Maxwell's equations 
imply the modulation of DC current by 
an external magnetic field in the static regime (SQUIDs) and the 
conversion of AC current into microwave radiation.

The ratio of the electromagnetic flux to
the flux quantum gives the Josephson length, the typical distance of 
variation of the phase $\lambda_J = { \Phi_0 \over  L J_c}$.
Depending whether their length $l$ is larger or not than $\lambda_J$, 
Josephson junctions exhibit significantly different behaviors. 
Small junctions for which $l < \lambda_J$ are used as SQUIDS to measure
magnetic fields in the static regime or as signal mixers where the 
steep IV characteristic at the gap is used to downshift an 
unknown high frequency signal by combining it with a known local
oscillator signal. 
In this latter context Salez et al \cite{Salez} suggested to 
use several unequally spaced junctions
, instead of just one, to get a better frequency response
over a large frequency band. Long Josephson junctions 
for which $l >> \lambda_J$ are described by the 1D
sine-Gordon equation
\be\label{sg1d}
\phi_{tt} - \phi_{xx} + \sin (\phi)=0 ,\ee
which admit the kink solutions varying from 0 to $2 \pi$. 
Long junctions have a characteristic dependence
of the maximum critical current in the presence of magnetic field 
because of the screening of the applied field by the supercurrent.
In the dynamical regime long junctions are used as microwave
generators but provide small output power and poor impedance
matching to the transmission line. To improve on this, series arrays
of junctions have been used, with success since they have been 
shown to synchronize
in the presence of external forcing \cite{bcsl99} \cite{vbsl02} giving
rise to an $N^2$ power output for $N$ junctions. Such behavior
has been understood both from the quantum point of view \cite{as01}
and the Resistive Shunted Junction (RSJ) equivalent circuit model 
\cite{fp02}. The cavity surrounding the junctions
stores radiation which can increase the output power of
the device. 

The mathematical description of such a system is a wave equation for
the cavity coupled to the 2D version of the sine-Gordon equation 
(\ref{sg1d}) via interface conditions.  
In \cite{bcf02,bc02} one of the authors 
studied this coupling for a long junction with a small surrounding passive
region. When the cavity is present 
only at the junction tip and the miss-match at the interface is
not too large, the kink crosses it unaffected and adapts its 
speed. When the passive region surrounds the junction, it sets the
kink speed. This ballistic kink motion will be absent for small
junctions so in the present work we consider the
coupling of a large cavity to one or a few small junctions. We want to
understand how junctions can talk to each other via the surrounding
passive region. Two approximations are assumed, we consider 
the 1D limit where transverse modes of the microstrip do not play a role
and neglect the phase variation inside the junctions, in other words
we neglect the junction length. This is valid for an array of short
junctions whose length remains smaller than the wave length of the
radiation.

This approach can be applied to high Tc superconductors where due to
a different symmetry of the wave function of the Cooper pairs one has
current phase relations that involve the higher harmonics \cite{lcti03}.
This different symmetry can also induce a negative coupling in (\ref{joseph})
leading to so-called $\pi$ junctions. Normal (0) and $\pi$ junctions can
be associated to form 0-$\pi$ junctions which have semifluxon solutions
that vary between 0 and $\pi$ and are attached to the discontinuity.
In the static regime the maximum DC current going through the system 
is strongly modulated for a short junction like for a SQUID while it
is constant for small field when the junction is longer \cite{l04}. The
IV characteristics give resonances that are similar to zero field steps
\cite{l04,gsgk04}. Note however the difference between the ballistic motion
of a kink in a (0) long Josephson junction and the hopping of semi-fluxons
in a (0-$\pi$) junction. As for 0 junctions, (0-$\pi$) junctions can
be organized as arrays \cite{r03,sg04} and these have interesting paramagnetic
behavior.

The model we introduce allows to describe accurately the coupling between 
small junctions in such an array. Both 0 and $\pi$ junctions can be considered
and disorder can be added. For simplicity we have considered the large
damping limit where it is possible to develop the analysis. We studied in
detail the cases of one and two junctions that remain fairly simple
and used our conclusions for the case of many junctions. For a single
junction, we found two limiting behaviors: the ohmic mode where the junction
acts as a pure resistor which stops waves and separates the cavity and 
the junction mode where the wave is homogeneous throughout the microstrip.
After introducing the model is section 2, we consider these limiting 
behaviors in sections 3,4 and 5.
These allow to understand in detail the features of the IV characteristics 
(section 6). The behavior of two junctions shown in section 7 
follows generally ohmic modes
and combined junction/ohmic modes. Using this, in section 8 we 
generalize the study to 
many junctions embedded in a microstrip and understand the influence of
the sign of the current density (0 or $\pi$ junctions) and external 
magnetic field. Conclusions are presented in section 9.

\section{The model}

\begin{figure}
\centerline{ \epsfig{file=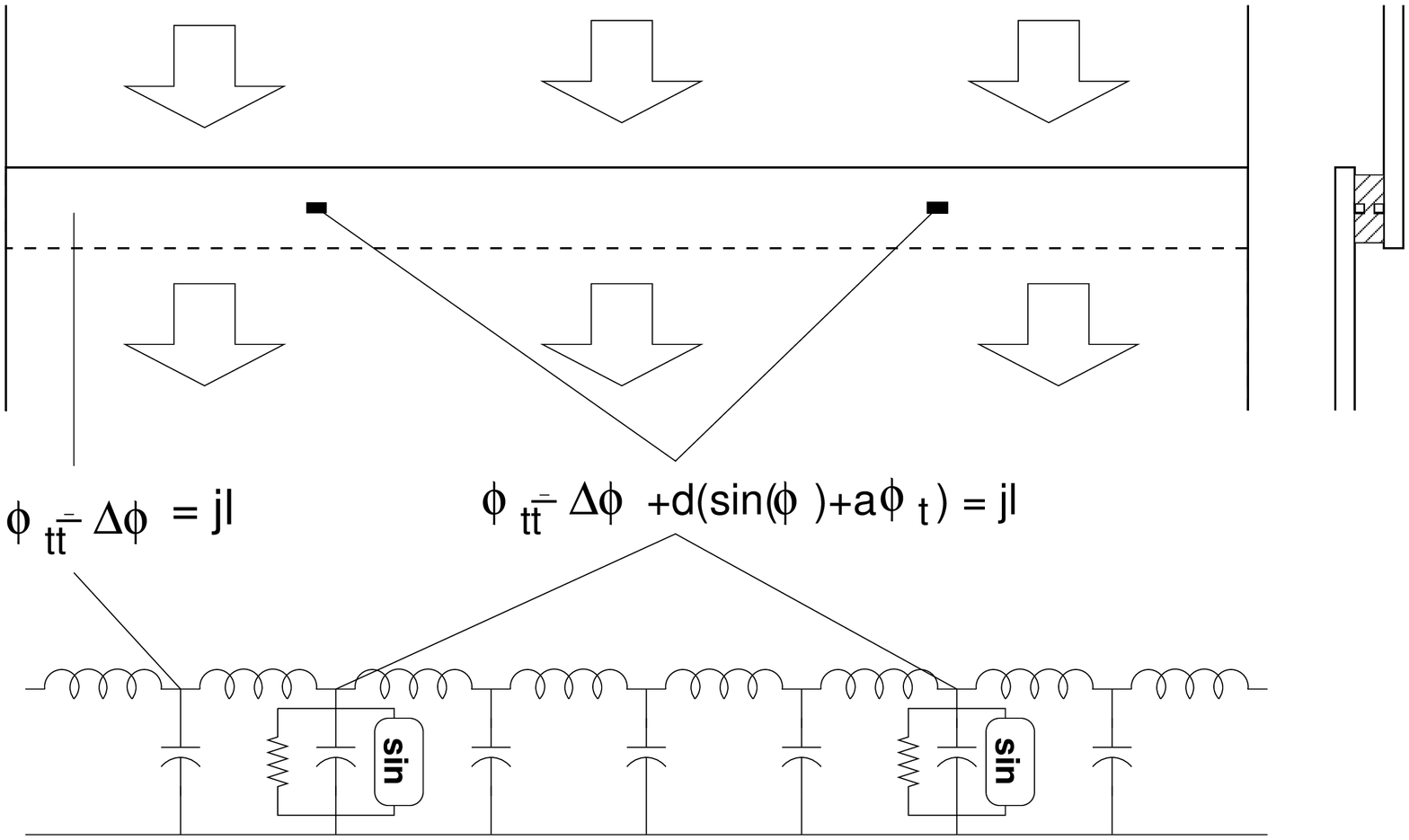,height=7 cm,width=10 cm,angle=0}}
\caption{Left top pannel, a schematic drawing of a narrow 
2D microstrip line containing
two small Josephson junctions, right top panel section at one of the junction. 
The bottom panel presents the equivalent 1D electric circuit.}
\label{f1}
\end{figure}

The device we model shown in the top panels of Fig. \ref{f1} (top view 
on the left and section on the right) is a narrow microstrip in
which are embedded a number of small junctions in the window
design. We proceed from the description of the two superconductors 
as two inductance arrays connected by capacitive elements (resp. and
a resistor and nonlinear Josephson element) in the passive (resp. junction)
region. The equivalent circuit is shown in the bottom panel of Fig. 
\ref{f1}. In the following subsections we show that the continuum limit
yields an inhomogeneous 2D sine-Gordon equation which we reduce to 1D
when the microstrip is narrow. For small junctions we then justify our delta
function approach (\ref{sgd}).

\subsection{RSJ model for distributed Josephson junctions}

A simple mathematical model of the window Josephson junction is to 
describe
each superconductor by an array of inductances $L$. The coupling 
elements
between two adjacent nodes in each array are, a capacitor $C$, 
resistance $R$
and Josephson current $I_c$ \cite{Likharev}. The Kirchoff laws at 
each couple of nodes $(i_b,i_t)$ in the bottom and top superconducting 
layers can be 
combined to give the relation expressing the conservation of currents 
at node $i$ in the device
\begin{equation}
\label{discr}
C_i \ddot \Phi_i + \sum_j  \frac {\Phi_i - \Phi_j}{L} + 
I_i^c \sin \frac {\Phi_i}
{\Phi_0} +  \frac {\dot \Phi_i}{R_i} = 0 \; ,
\end{equation}
where $\Phi = \Phi_t - \Phi_b$ (resp. $\Psi = \Psi_t - \Psi_b$), 
is the phase difference between the two superconductors 
in the junction (resp. passive) part,  the summation $\sum_j$ is 
applied to 
the nearest neighbors, $I_i^c$ is the critical current which is non 
zero
only in the junctions and $R_i$ is the resistance associated to the
current of quasi-particles again non zero
only in the junctions. In agreement with experiments, we have 
assumed the same surface inductance for
the whole sample but have taken into account the variation of the 
capacity
$C$ and resistance $R$ from the junctions to the rest of the cavity.
Note that equation (\ref{discr}) is a discretisation of Maxwell's 
equations 
(wave equation part) and Josephson constitutive equations (sine term), 
assuming an electric field normal to the plates, a magnetic field in 
the 
junction plane and perfect symmetry
between the top and bottom superconducting layers. We can now obtain 
the model
in the continuum limit, more suitable for analysis.

\subsection{Continuum limit}

The continuum version of the system (\ref{discr})
can be derived by introducing the following quantities
per  unit area ($a^2$) of elementary cells 
of length $a$.
\begin{equation}
\label{normcjr}
\overline {C_{J}}=\frac {C_{J}} {a^2},~~~j_i^c= \frac {I_i^c} {a^2},~~~
\overline{r_i} = R_i a^2.~~~
\end{equation}
We normalize the phases by the flux quantum $\Phi_0$,
\begin{equation}
\label{normphi}
\varphi_i=\frac {\Phi_i}{\Phi_0} 
\end{equation}
and introduce the Josephson characteristic length 
\be \label{lambdaj}
\lambda_J^2=\frac {\Phi_0}{j_c L_J} .
\ee
Notice that in this 2D problem the inductance associated to each cell
is equal to the branch inductance $L$. 
To differentiate the junction regions from the passive regions we
introduce the indicator function $g_i$ such that $g_i=1$ in the
junctions and $0$ elsewhere. We can then write
\be\label{indic}
j_i^c =  g_i j^c~~,~~
{1\over \overline{r_i}}=g_i{1\over \overline{r}}
~~{\rm and} ~~ 
{\bar C_i}= g_i{\bar C_J}  +(1-g_i){\bar C_I}
\ee
where $j^c$, $\overline{r}$, ${\bar C_J}$ and ${\bar C_I}$ 
are respectively the critical current densities, the quasi 
particle resistance in the junction, the capacitance per unit area in
the junction and the capacitance per unit area in
the passive region.

We  substitute relations (\ref{normcjr}) - (\ref{indic}) in 
(\ref{discr}) 
and obtain 
\begin{equation} \label{discre2}
L {\overline C_I} \ddot \varphi_i 
+  \sum_j \frac {\varphi_i - \varphi_j}{a^2}
+ g_i \left [ L({\overline C_J}-{\overline C_I}) \ddot \varphi_i 
+ {1 \over \lambda_J^2} \sin \varphi_i  
+\frac {L}{\overline{r}} \dot \varphi_i \right ] = 0 \; ,
\end{equation}
which in the continuum limit yields
\begin{equation} \label{cont1}
L{\overline C_I} \varphi_{tt}
-\Delta \varphi 
+ {1 \over \lambda_J^2}g(x,y) \left [
L \lambda_J^2 ({\overline C_J}-{\overline C_I}) \varphi_{tt}
+ \sin \varphi 
+ \frac {L \lambda_J^2}{\overline{r}} \varphi_t \right ]=0 . 
\end{equation}
To obtain the final equation we rescale time by the inverse of the
plasma frequency $\omega_I^{-1}= \lambda_J \sqrt{ L {\overline C_I}} $
so that $\tilde t = \omega_I t$. As unit of space we use the
Josephson length, 
$ \tilde x = x/\lambda_J,~\tilde y =y/\lambda_J $. 
The final equation is then 
\be
\label{insg}
\varphi_{tt} -\Delta \varphi
+ g(x, y)( c \varphi_{tt} + \sin \varphi + \alpha \varphi_{t})=0 ~,
\ee
where the
tildes have been omitted for simplicity
and where the coefficients $c$ and $\alpha$ are
\be\label{calpha} 
c = {{\overline C_J}\over{\overline C_I}} -1~~,~~ 
\alpha = {\lambda_J \over {\bar r}  }\sqrt{L \over {\overline C_I}}\ee 

The boundary conditions represent the input of an external
current $I$ or magnetic field $b$ on the device. They are
\be\label{overlap}
\varphi_x|_{x=0}=b -(1-\nu){I\over 2 w}~~,~~ 
\varphi_x|_{x=l}=b +(1-\nu){I\over 2 w}~~,~~
\varphi_y|_{y=0}= -\nu {I\over 2l}~~,~~
\varphi_y|_{y=w}= \nu {I\over 2l},\ee
where $0<\nu<1$. The case $\nu=1$ corresponds to a pure 
inline current feed and $\nu=0$ to a pure overlap current feed.
We will mostly consider the latter case in the discussion.

\subsection{Reduction to a 1D problem}

When the microstrip is narrow and the junctions occupy the
whole width we can assume that $g$ is independent of $y$.
We can then write 
\be \varphi(x,y,t) = \nu I {y^2 \over 2s} 
+ \sum^\infty_{n=0} \phi_n(x,t) \cos \left({n \pi y \over  w}\right), 
\ee
where the first term takes care of the current feed and the second
satisfies homogeneous Neumann boundary conditions for $y=0,w$.

The calculations detailed in \cite{bcf02} show that for
a narrow width and small current only the first mode $\phi_0$
needs to be taken into account. We then obtain the following
equation for $\phi_0$ where the 0 has been dropped for
simplicity
\be\label{final1d}
{\phi}_{tt} -{\phi}_{xx}
+ g(x)( c {\phi}_{tt} + \sin {\phi} + \alpha {\phi}_{t})=\nu{I \over s} 
~,
\ee
with the boundary conditions 
${{\phi}_x}_{x=0}= -(1-\nu){I l \over 2 s}$  ~~
${{\phi}_x}_{x=l}= (1-\nu){I l \over 2 s}$.  
In the following we will write
\be\label{curdens}
j = {I \over s},\ee
and assume an overlap current feed ($\nu=1$) except in section 7.

\subsection{The delta function model}

As the area of the junction is reduced the total super-current is
reduced and tends to zero. To describe situations where the area 
is small so that the variations of the phase can be neglected in the 
junction but the supercurrent is significant we introduce the
following delta function model.
Introduce the function $g(x) = d_a g_h(x)$\\
$g_h(x) = \left\{
\begin{array}{l r}
1/(2h) & a-h<x<a+h \\
0 & {\rm elsewhere} \\
\end{array} \right.$

\begin{equation} \label{sggh}
\phi_{tt}- \phi_{xx} + d_a g_h(x) \left(\sin(\phi)
+ \alpha\phi_t + c \phi_{tt} \right) =j.
\end{equation}
In the following we assume $\phi$ to be $C^2$ in $x$ and $t$ and
$\phi_{ttx}$ to be continuous.
We now show that equation (\ref{sggh}) converges when
$h\rightarrow 0$ to
\begin{equation} \label{sg2}
\phi_{tt}- \phi_{xx} + d_a \delta(x-a)\left(\sin(\phi)
+ \alpha\phi_t + c \phi_{tt} \right) = j.
\end{equation}
For that introduce the Taylor expansion of $\phi$ near $x=a$
\begin{equation}\label{taylor}
\phi(x,t)=\phi(a,t) + (x-a)\phi_x(a,t) + O((x-a)^2) .
\end{equation}
In the following for simplicity of writing we will omit the
$t$ and $x$ dependencies. We integrate (\ref{sggh}) from $a-h$ to $a+h$
(see details in appendix: Continuum limit). The
first term 
$$\int_{a-h}^{a+h} \phi_{tt} dx \rightarrow O ~~{\rm when} ~~h 
\rightarrow 0, $$
the second one
$$\int_{a-h}^{a+h} \phi_{xx} dx=\phi_x(a+h)-\phi_x(a),$$
and the third one
$$\lim_{h\rightarrow 0}{d_a \over 2 h}\int_{a-h}^{a+h}
(\sin(\phi)+ \alpha\phi_t + c \phi_{tt})dx = d_a
\left( \sin(\phi(a))+ \alpha\phi_t(a) + c \phi_{tt} (a) \right),$$
when $h \rightarrow 0$ which is consistent with the model
\begin{equation} \label{sgcd}
\phi_{tt}- \phi_{xx} + d_a \delta(x-a) \left( \sin(\phi) + \alpha\phi_t 
+
c \phi_{tt} \right)= j,
\end{equation}
together with the boundary conditions 
$\phi_x\left (x=0\right)=0~,~~~
\phi_x\left (x=l,t\right)=0.$

In the following we will consider the case $c=0$ for simplicity. We 
assume the
same capacity in the junction and in the passive region
so that the final equation is
\be\label{sgd}
\phi_{tt}- \phi_{xx} + d_a \delta(x-a) \left( \sin(\phi) + \alpha\phi_t 
\right)= j.
\ee
This equation can be considered as a 1D model of a point Josephson
junction embedded in a microstrip cavity.

\section{The single junction case}

The partial differential equation (\ref{sgd}) involves a distribution, 
however
its solution is differentiable except at the junction points.
To solve it numerically we integrate the equation
over reference intervals using the finite volume approach.
The resulting system of ordinary differential equations is then solved
using the DOPRI5 4th and 5th order Runge-Kutta variable step 
integrator developed by Hairer and Norsett \cite{hairer}.
Details of the numerical code are given in the second part of the appendix.

We now consider the two limiting cases of a static solution and
high voltage for which simple expressions of
the solution can be obtained. In particular these have been used to 
validate our numerical procedure.

\subsection {Static behavior}

We assume $\phi_t = \phi_{tt} \equiv 0$ in (\ref{sgd}) and get
\begin{eqnarray}
\label{statique}
-\phi_{xx} + \delta (x-a) d_a \sin (\phi) = j \\
\phi_x|_{x=0,l} =0
\end{eqnarray}
For $x\neq a$, $-\phi_{xx}=j $ so that for
$x<a$, $\phi= \phi_l \equiv -\frac{j}{2}x^2 + C_l x + D_l$ and for
$x>a$, $\phi= \phi_r \equiv -\frac{j}{2}x^2 + C_r x + D_r$.
Integrating (\ref{statique}) over the whole domain we get the
conservation of current
\be\label{conscur} d_a \sin(\phi(a))= j l,\ee
so that no solution exists for $|jl| > d_a$.

The solution is obtained as usual by assuming continuity of the 
phase and using the jump condition on the gradient obtained 
by integrating on a small domain including
the junction
\be\label{jump}[\phi_x]_{a^-}^{a^+}=d_a \sin(\phi(a)).\ee
Using these two constraints we obtain the static solution
\begin{equation}
\phi_s(x)=  \left\{\begin{array}{lr}
\frac{j}{2}(a - x)(a + x) + \arcsin \left ({j l \over d_a}\right) & 
x<a\\
\frac{j}{2}((a - x)(x+a+2l) + \arcsin\left({j l \over d_a} \right) & 
a<x
\end{array}\right .\end{equation}

\subsection{High voltage behavior}

Since the Josephson current is bounded, its influence in (\ref{sgd})
will decrease as the current $j $ is increased. In this limit
of high current the phase is growing very fast so that $\phi_t$
is close to the average voltage $V\equiv \left<\phi_t\right>_t$.
We can write approximately $\phi(x,t)= V t +\phi_v(x)$ so that
we can average (\ref{sgd}) assuming $\left<\phi_t(x,t)\right>_t = V$, 
$\left<\sin(\phi(a))\right>_t=0$ and $\left<\phi_{tt}\right>_t=0$, 
yielding
\be \label{hvolt}
-{\phi_v}{xx} + \delta\left(x-a\right)d_a \alpha V= j ,\ee
with the boundary conditions ${\phi_v}_x|_{x=0,l}=0$.
Integrating this equation over the whole domain 
we get $V= {j l \over d_a \alpha}$.

Using the same constraints as in the static case (continuity of the
phase and jump of its gradient at the junction) we get up to
a constant the high voltage solution $\phi_v$
\begin{equation}
\phi_v(x,t)= \left\{\begin{array}{lr}
{j\over 2 }(a-x)(a+x) + Vt + C & x<a \\
{j\over 2 }(a-x)(a+x-2l)+ Vt + C & a<x
\end{array}\right .\end{equation}
Notice that this and precedent solutions have the same spatial 
behavior;
$\phi_s-\phi_v = Vt + C$. We will not find this result with more 
junctions
(except in particulary cases).

\subsection{Fourier representation }

Because of the homogeneous Neumann boundary conditions one can write the
solution as a cosine Fourier series
\begin{equation}\label{four}
\phi(x,t) = \sum_{n=0}^{+\infty} A_n(t)\cos\left(\frac{n\pi x}{l} 
\right).
\end{equation}
The solution of (\ref{sgd}) has a discontinuous first spatial
derivative to that the amplitude of the mode $A_n$ decreases as $1/n^2$ 
\cite{carslaw}. This means that in general many modes are needed 
for the description of the solution. However since the system is
close to a linear one we will see that these modes provide insight into
the limiting behaviors.

 Plugging (\ref{four}) into (\ref{sgd}) and projecting we get the
evolution of the modes, 
\begin{equation}\label{a0}
l {A_0}_{tt} + d_a (\sin(\phi_a) + \alpha \phi_{at})
-j l =0 ,\end{equation}
\begin{equation}\label{an}
{A_n}_{tt} + \left({n \pi  \over l}\right)^2  A_n
+ {2 d_a \over l } c_n ( \sin(\phi_a) + \alpha \phi_{at})
=0 ,  \end{equation}
where we have introduced the coupling coefficients 
associated to the junction at $x=a$,
$c_n=\cos\left(\frac{n\pi a}{l} \right)$.
Equations (\ref{a0}) and (\ref{an}) are coupled through the
term $\sin(\phi_a) + \alpha {\phi_a}_t$.  Due to this particular
feature, two limiting behaviors can be expected.
\begin{itemize}
\item The ohmic mode for which
\be\label{omode} 
d_a  \alpha \phi_a|_t - j l=0 ~.\ee
Then the equations for the $A_n$ have a right hand side so that
the cavity modes are driven by the junction.

\item The junction mode such that
\be \label{jmode} 
d_a ( \sin(\phi_a) + \alpha {\phi_a}_t)  -j l =0 ~.\ee
In this case the equations for the $A_n$ have a constant right hand 
side 
($= j l$). We will show that in this case the cavity is synchronized 
with the junction.
\end{itemize}

The equations giving the evolution of $\phi_a$ in these two 
limiting behaviors can be solved yielding for the ohmic mode
\be\label{som}
\phi_a = { j l  \over d_a \alpha} t + C_1 ~, \ee
where $C_1$ is a constant and a voltage
\be\label{vohm}
V_o = { j l  \over d_a \alpha}.\ee
For the junction mode the
equation can be solved using separation of
variables and we get 
\be \label{sjm} \phi_a =2\arctan\left\{ 
{d_a \over j I } \left[ \alpha V_j    \tan
\left( {V_j \over 2 } \left(t+C_2\right)
 \right) +1 \right] \right\}~~,\ee
where the junction mode voltage $V_j$ is 
 \be\label{vjon}
 V_j = {1 \over \alpha}\sqrt{\left( {j l \over d_a}\right)^2 -1 }\ee
where $C_2$ is a constant.

\begin{figure} 
\centerline{\epsfig{file=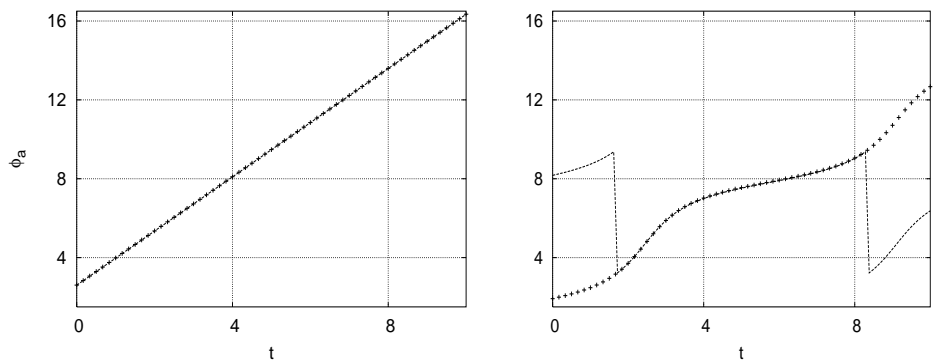,height=14 cm,width=5.5cm,angle=270}}
\caption{plot of $\phi_a(t)$ for $j = 0.137414$ for a junction position
$a=3.43$, $d_a=1$ and length $l=10$. The left panel shows 
the ohmic mode regime and the
right panel the junction mode regime. The numerical solution of
(\ref{sgd}) is shown with the crosses while the analytical estimates
(\ref{som}) and (\ref{sjm}) are in dashed line. }
\label{f2}
\end{figure}

Fig. \ref{f2} shows these two limiting behaviors for a junction
placed in $a=-1.57$ with $d_a=1$ $l=10$ and a current $j\approx 
0.1374$.
The numerical solution is presented with the crosses while 
the analytical estimates (\ref{som}) and (\ref{sjm}) are shown 
in continuous line.
The left panel corresponds to the ohmic mode and one can
see the excellent agreement of the numerical solution with the
line given by (\ref{som}). The right panel presents
the junction mode and the extension of the estimate (see below), again 
the
agreement is very good. 
Notice that for the same time interval the variation of $\phi_t(a,t)$ 
is different for the two panels of Fig. \ref{f2} so that we will get 
a different average voltage for these two behaviors.

\section{The ohmic mode}

In this case $\phi_t(a,t)=V_o={j l \over \alpha d_a}$. To get
the field we need to solve on each subdomain $[0,a[$ and $ ]a,l]$
the following problems
$$\begin{array}{cc}
\left\{\begin{array}{l}
\phi_{tt}-\phi_{xx}= j  \\ 
\left .\phi_{x}\right|_{0}=0,~
\left .\phi_{t}\right|_{a}=\frac{j l }{\alpha d_a}
\end{array}\right.&
\left\{\begin{array}{l}
\label{onde1}
\phi_{tt}-\phi_{xx}= j \\ 
\left .\phi_{t}\right|_{a}=\frac{j l }{\alpha d_a},~
\left .\phi_{x}\right|_{l}=0 .
\end{array}\right.\\
\end{array}$$
Consider the left subdomain $[0,a[$, we introduce 
$\phi\left(x,t\right)=v\left(x,t\right)+\frac{ j l t}{\alpha d_a}$
so that if $\phi$ solves the left problem then $v$ solves
\begin{eqnarray}
\label{onde2}
v_{tt}-v_{xx}= j \\
\left. v_x \right |_{0}=0~\left. v_t\right|_{a}=0
\end{eqnarray}
A particular solution of (\ref{onde2}) is $v_l=
-\frac{j}{2}x^2 + D_l$ where
$D_l$ is an integration constant.
Then $w= v -v_l$ solves the homogeneous wave equation 
$$\partial_{tt} w-\partial_{xx} w=0 , $$ 
with $w_x|_0=0$ and $w_t|_a=0$.
We can assume without loss of generality that
$w\left(x,0\right)=0$.
The condition $w_x|_0 =0$ imposes a solution
of the type 
\begin{equation}
w\left(x,t\right)=a_l\sin\left(V_l t\right)
\cos\left(V_l x\right) ,
\end{equation}
where $V_l$ is a constant. Also, $w_t|_{a}=0$ imposes a
condition on $V_l$ because 
$$ w_t\left(x=a,t\right) =a_l \sin\left(V_l t\right) \cos(V_l a)=0$$ 
is possible for all $t$'s only if  $\cos\left(V_l a\right)=0$. So 
that we get $$V_l={ 2k_l+1 \over 2 }\frac{\pi}{a}.$$
The problem is identical for $x \in \left]a,l\right]$
and using similar arguments we 
get $v_r=-\frac{j}{2}\left(x-l\right)^2 + D_r$ and
$$V_r={2k_r+1 \over 2} \frac{\pi}{a-l} .$$
We obtain a solution for the equation (\ref{sgd}).
The phase on the whole domain is then
\begin{equation}
\label{totalcos}
\phi\left(x,t\right) =  \left\{ \begin{array}{l r}
\frac{j l t}{\alpha d_a} + a_l \sin\left(V_l t\right)\cos\left(V_l 
x\right) + \frac{j}{2} (a-x)(a+x)  & x<a \\
\frac{j l t}{\alpha d_a} + a_r \sin\left(V_r t\right)\cos\left(V_r
\left(x-l \right) \right) + \frac{j}{2} (a-x)(a+x-2l) & a<x
\end{array} \right. \end{equation}
From the problem (\ref{sgd}) integrating the equation on a small 
interval
centered on $x=a$ and using (\ref{omode}) we get the jump condition 
$$[\phi_x]_{a^-}^{a^+} -j l
= d_a\sin(\phi(a,t)) = d_a\sin (\frac{j l t}{\alpha d_a}).$$
On the other hand from (\ref{totalcos}) we get 
$$\begin{array}{ccl}
\left[\phi_x\right]_{a^-}^{a^+} -j l &=&  a_l 
\sin\left(V_l t\right)
\sin \left( V_l a\right)-a_r V_r \sin\left(V_r t\right)
\sin ( V_r\left(a-l\right)) \\
  & \equiv & C_1 \sin\left(V_l t\right)-C_2 \sin\left(V_r t\right)\\
\end{array}$$
Since $C_1$ and $C_2$ are independent of $t$, this implies that 
$\sin(V_l t)$, $\sin(V_r t)$ and  $\sin({j l t\over \alpha d_a})$ have 
the 
same period. We obtain, 
\be \label{VleftVright}
{ 2k_l+1 \over 2 }\frac{\pi}{a}= {j l \over \alpha d_a}=
{2k_r+1 \over 2} \frac{\pi}{a-l}=V,\ee
if $a_l$ and $a_r$ are not equal to zero. Then one can write the
solution as
\be\label{leftaright}
\phi(x,t) =  \left\{ \begin{array}{l r}
\phi_v(x,t) + a_l \sin\left(V t\right)\cos\left(V x\right) & x<a \\
\phi_v(x,t) + a_r \sin\left(V t\right)\cos\left(V \left(x-l \right) 
\right) & a<x.
\end{array}\right .\ee
where $ \phi_v$ is the high voltage solution. Because of this the 
instantaneous
voltage $\phi_t$ has a discontinuous first derivative at $x=a$.

Assuming that $\alpha=1$ we see that the consistency equation 
(\ref{VleftVright}) 
implies 
$$(a-l) 2 k_l - 2a k_r = l.$$
There are very few solutions for $a$ rational et $l$ integer and for 
most $a$ there are none. Let us look at other possibilities. Notice 
first
that space homogeneous functions ($a_l=0$ or $a_r=0$) are solutions. 
Having said this
it is easy to see that one can obtain left only or right only 
oscillations. \\
left only: $\phi(x,t) =  \left\{ \begin{array}{l r}
\phi_v(x,t) + a_l \sin(V t)\cos(V x) & x<a \\
\phi_v(x,t) & a<x
\end{array}   \right .$\\
right only:
$\phi(x,t) =  \left\{ \begin{array}{l r}
\phi_v(x,t) & x<a \\
\phi_v(x,t) + a_r \sin(V t)\cos\left(V\left(x-l \right)\right) & a<x
\end{array}   \right .$\\
Again the instantaneous voltage at the junction has a discontinuous
first derivative.

\begin{figure} 
\centerline{\epsfig{file=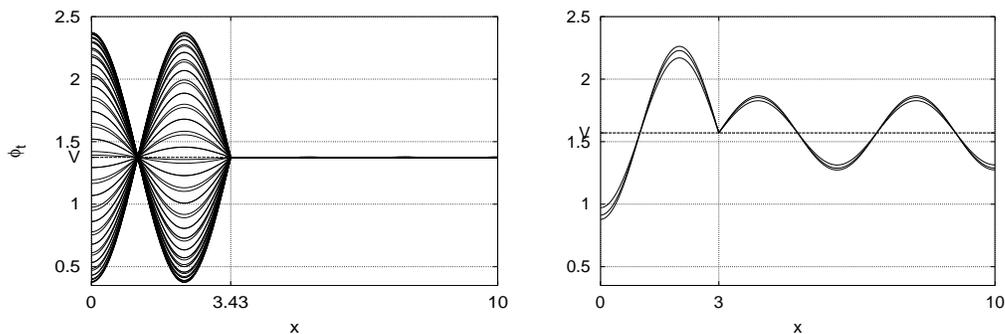,height=15 cm,width=5cm,angle=270}}
\caption{Instantaneous voltage $\phi_t\left(x,t\right)$
as a function of $x$ for different times for the ohmic mode: left
only oscillation for the junction shown in Fig. \ref{f2} placed at 
$a=3.43$ (left panel) and left and right oscillation for a junction placed at 
$a=2$ (right panel). The average voltage is indicated in both figures, $V=j l 
=1.37$ in the left panel (same as in Fig. \ref{f2})
and $V=\frac{\pi}{2}\left(=j l \right)$ on the right panel. 
The other parameters are $\alpha=1$, and $d_a=1$.}
\label{f3}
\end{figure}

Fig. \ref{f3} shows the instantaneous voltage as a function of $x$ for
different times for two such ohmic modes, the left only oscillation 
(left panel)
and right and left oscillations (right panel). The left panel shows
$\phi_t(x,t)$ for 60 values of time equidistributed between 0 and 10.
The parameters are the same as for Fig. \ref{f2}. In this case we have 
$k_l=1$
so that about 1.5 periods exist to the left of $x=a$. Notice how the 
amplitude
of $\phi_t$ is practically constant and equal to the average voltage
$V=j l =1.37$ on the right of $x=a$. The right panel corresponds to a 
right and left oscillation for $a=2$ and we show three consecutives 
times, 
$t=0, 0.125$ and $0.250$. Here ${ 2 k_l+1 \over 2 k_r +1} = {a \over 
l-a}=3/7$
so that we get about 1 period to the left of $x=a$ and two periods to
the right of $x=a$. We show for comparison
$V=\frac{\pi}{2}\left(=j l \right)$ the average voltage. Notice in both 
cases
the discontinuity of $\phi_t$ at $x=a$.

Let us show that this is consistent with
the whole equation (\ref{sgd}), for the intensity $I=V/ \alpha d_a$.
Consider the left only solution
\begin{equation}
\label{gauchecos}
\phi\left(x,t\right)=  \left \{ \begin{array}{l r}
Vt+ \frac{a_l}{V}\sin(V t)\cos(Vx)-\frac{j}{2}[x^2-a^2]& x < a \\
Vt-\frac{j}{2}\left[(x-l)^2+(a-l)^2\right]& x > a
\end{array}   \right .
\end{equation}
with, $V=\frac{2k_l+1}{2}\frac{\pi}{a}$.
On the interval $[0,a[$ (respectively $]a,l]$) $\phi$ satisfies
the linear wave equation 
$\phi_{tt}-\phi_{xx} = j$ with the boundary condition
$\phi_x|_{0}=0$ (respectively $\phi_x|_{l}=0$).
We now integrate (\ref{sgd}) on $[0,l]$ to obtain
\be
\int_{0}^a \phi_{tt}dx+\int_a^{l}\phi_{tt}dx +
\int_{0}^{l}\phi_{xx}dx+d_a\sin\left(Vt\right)+\alpha d_a V = jl, \ee
so that
$$-a_l \sin(V t)\left[\sin(Vx)\right]_0^a + 0 + 0 + 
d_a\sin\left(V t\right) = 0.$$
The end result is 
$$(-1)^{k_l+1} a_l \sin(V t)=d_a\sin(V t)$$
which implies $a_l=(-1)^{k_l+1} d_a$ and the equation is
balanced.

\section{The junction mode}

The expression (\ref{sjm}) a priori defined only for a given time 
interval
can be $C^{\infty}$ extended for all times using the continuation
$$f \left(t+k\frac{2\alpha \pi}{\sqrt{(j/d_a)^2-1}}  \right)= 
f \left(t\right)+k 2\pi ~,$$
where $k$ is an integer.
The pseudo-period $T$ of $\phi_a$ is $T=\frac{2\alpha 
\pi}{\sqrt{(j/d_a)^2-1}}$
so the voltage is $V=\frac{2\pi}{T}={1 \over 
\alpha}\sqrt{(j/d_a)^2-1}$. 
This can also be written
\be\label{ji}
j = {d_a \over l} \sqrt{\left(\alpha V\right)^2+1}\ee
So we deduce from $A^{''}_0(t)=0$ and the voltage expression 
(\ref{vjon}) 
that $A_0(t)=Vt+C$.

\begin{figure} 
\centerline{\epsfig{file=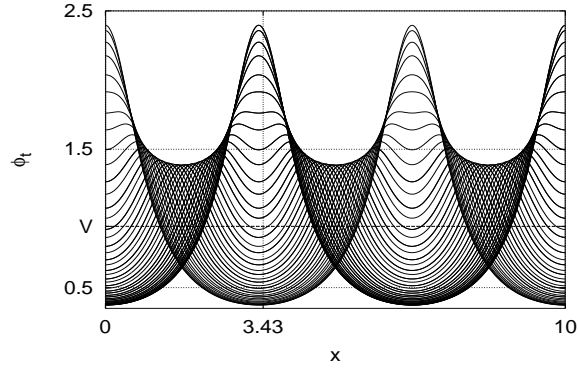,height=8 cm,width=5cm,angle=270}}
\caption{Instantaneous voltage $\phi_t(x,t)$ in the
junction mode as
a function of $x$ for 60 values of time equidistributed between 0 and 
10.
The parameters are the same as for Fig. 1. The average voltage $V$ is
indicated on the vertical axis and the junction's position is indicated
on horizontal axis.}
\label{f4}
\end{figure}

Let us now calculate the contribution of the
Fourier modes and 
give a solution to the whole equation (\ref{an}). From (\ref{an})
and the previous equality, we obtain: 
\be \label{a_pjunc}
A_p^{''}\left(t\right) +  \left(\frac{p \pi}{l}  \right)^2
A_p\left(t\right) = - 2 j c_p
\ee
The solution of this equation, $A_p$ is:
\be \label{solA_pjonc}
A_p \left(t\right)= \gamma_p \cos \left(\frac{p \pi}{l}t \right) 
+ \beta_p\sin \left(\frac{p \pi}{l}t \right)
-\frac{2 j l^2}{(p \pi)^2} c_p
\ee
Now we have the expression of all the Fourier modes. 
We know that $\phi(a,t) = \sum_{n=0}^{+\infty}A_n(t) c_n^a$.
Substituting the Fourier modes (\ref{solA_pjonc}) into 
this expression we obtain:
\be \label{pjmi} 
\phi(a,t)=Vt+C+\sum_{p=1}^{+\infty}\left[\gamma_p \cos \left(\frac{p 
\pi}{l}t \right) 
+ \beta_p \sin \left(\frac{p \pi}{l}t \right)-\frac{2 j l^2}
{(p \pi)^2} c_p\right]c_p ,
\ee 
The series in the last term is uniformly convergent so that the term 
can be
collected in the form of a constant yielding the final simpler 
expression
\begin{equation}\label{pjm} \phi(a,t)=Vt+C_2+\sum_{p=1}^{+\infty}
\left[\gamma_p c_p\cos \left(\frac{p \pi}{l}t \right) 
+ \beta_p c_p \sin \left(\frac{p \pi}{l}t \right)\right]
\end{equation}
Introduce $\phi(a,t)-Vt+C_2 = u(t)$. We can choose $C_2$ as $\int_0^T 
u(t) dt =0$.
Notice that $u$ is a $C^{\infty}$ and $T$-periodic function. 
We can make Fourier projection on it. 
$$u(t) = \sum_{p=1}^{+\infty}\left[\gamma_p c_p\cos \left(\frac{p 
\pi}{l}t \right)
+ \beta_p c_p \sin \left(\frac{p \pi}{l}t \right)\right] $$
Until $c_p \neq 0$, we have
\begin{eqnarray}
\gamma_p = \frac{2}{c_p T}\int_0^T u(t) \cos\left(\frac{p 
\pi}{l}t\right) dt \\
\beta_p = \frac{2}{c_p T}\int_0^T u(t) \sin\left(\frac{p 
\pi}{l}t\right) dt 
\end{eqnarray}
We can make the following remarks 
\begin{enumerate} 
\item $V=\frac{2\pi}{T}=\frac{n\pi}{l}$ so $\frac{1}{T}=\frac{n}{2l}$ 
and $u$ is $\frac{2l}{n}$ periodic.
\item If $V=\frac{n\pi}{l}$ the terms of the Fourier series that aren't
zero are the multiples of $n$.
\item If $c_n=0$ then the junction modes for $V=\frac{n\pi}{l}$ can't 
exist.
For example when the junction is in the middle of the circuit, then 
$c_n^a=0$
for $n=2k+1$ and junction modes exist only for $V=\frac{2k\pi}{l}$ 
where $k$ is 
an integer (see left panel of figure \ref{f5}).
\item When $|c_n|$ is very small, the circuit in the junction mode 
stores
a lot of energy but is unstable. We will see this in the next section.
\end{enumerate}
Now remember (\ref{four}), we can give the explicit solution of the 
phase 
in the $n$ th junction mode, up to a constant
\begin{equation} \label{modjonc}
\phi(x,t) = {n \pi \over l} t + \sum_{p=1}^{+\infty} 
\left [\gamma_p \cos \left(\frac{p n \pi}{l}t\right) + 
\beta_p\sin \left(\frac{p n \pi}{l}t \right) \right ]
\cos\left(\frac{p n\pi}{l} x \right) ~.
\end{equation}
Notice that the instantaneous voltage $\phi_t$ has a continuous 
derivative at the junction position $x=a$.
Fig.  \ref{f4} shows the instantaneous voltage $\phi_t(x,t)$ as a 
function of $x$ for 60 consecutive times for the same parameters as
in Fig. \ref{f2}. Notice how the solution oscillates over the whole
domain and that one cannot see the junction position. This type of
motion is similar to the transparency observed in the reflection
coefficient of a stripe \cite{lamb} when an integer number of 
half-periods
of the wave "fits" in the stripe.

\section{Single junction case: IV characteristics}

For a large enough time the system finds a stationary state where the
energy provided by the input current is balanced by dissipation. This 
state 
is described by the IV characteristic curve. This is measured
on real devices and it is therefore very important to understand its 
features. 

To compute it we fix the current and run the system for about 
$t_1=3000$ 
time units so that
all transients have died off. Then we compute the average voltage
\be\label{avolt}
\left<\phi_t(a,t)\right>_t\equiv 
{\phi(a,t_1+T)-\phi(a,t_1)\over T}\ee 
at the junction. The value $T$ is about 2000. The space average of the 
average voltage in time 
$\left<\frac{1}{l}\int_{0}^{l} \phi_t (x,t) dx\right>_t$
is compared to this quantity to ensure that the
system is completely thermalized. Only then is the voltage recorded.

The IV curves are computed for 99 steps of increasing current $j$ from 
0 to 4
starting with an initial condition $\phi(x,0) \equiv 0$ and 
$\phi_t(x,0) \equiv 0$. We then decrease the current from 4 to 0.
All plots show these two curves for increasing and decreasing current.

\subsection{Bounds for IV curves}

\begin{figure}
\centerline{\epsfig{file=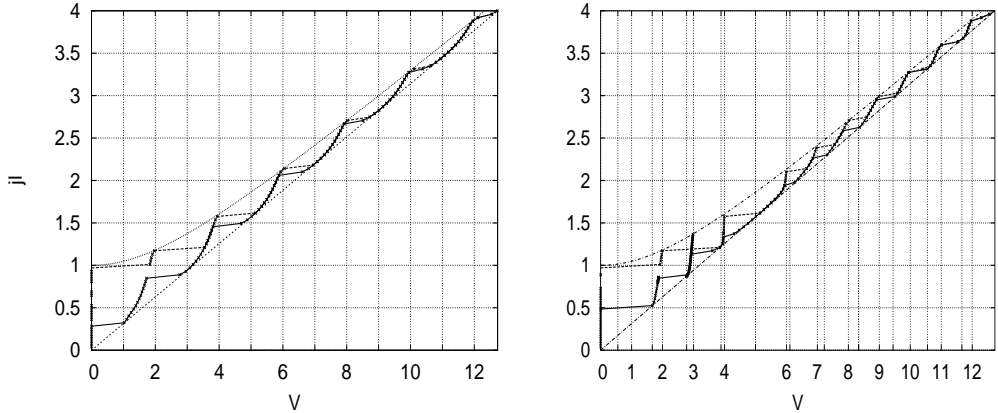,height=15. cm,width=6cm,angle=270}}
\caption{Two IV characteristic for a single junction placed in the
center of the microstrip at $a=5$
(left panel) and near the right edge at $a=9$ (right panel).
Only the positions of the junction modes are indicated in units of
$\frac{\pi}{10}$. The other ticks correspond to the ohmic modes.
For each panel two curves are presented, the top one is for increasing 
current
while the bottom one is for a decreasing current showing a clear 
hysteresis.
The minimum value of the current is 0, the maximum is 4 and
the stepping is $\frac{4}{99}$. The other parameters are
$d_a=\alpha=1$.
}
\label{f5}
\end{figure}

We have isolated two characteristic behaviors of the circuit. These 
allow
us to bound the IV curves for a single junction in a
microstrip.  Fig. \ref{f5} shows two IV curves for 
a centered junction $a=5$ (left panel) and an off-centered
junction $a=9$ in a microstrip of total length $l=10$. 
For each case we plot the limiting behaviors given by the
ohmic mode and junction mode voltages (\ref{vohm}) and (\ref{vjon}).
$V, j$ are bounded by the relations 
$$ d_a \alpha V \leq jl \leq d_a \sqrt{(\alpha V)^2 + 1}~,~{\rm or}~~
{1 \over \alpha}\sqrt{\left({jl \over d_a}\right)^2-1}\leq V \leq 
{jl \over d_a \alpha}$$
The ohmic modes, (resp. the junction modes) only occur
for voltages such that $V=V_l=\frac{2 k_l +1}{2}\frac{\pi}{a}$ or
$V=V_r = \frac{2 k_r + 1}{2}\frac{\pi}{a-l}$ (resp.  $V={k\pi \over l}$).
The voltage (horizontal) axis is labeled with the indices corresponding
in each case to the junction modes. The ohmic modes are marked with
ticks but not labelled. 
We always go from an ohmic mode 
to a junction mode and vice versa. As the current is increased the 
junction mode ceases to exist and the
system then jumps to the closest ohmic mode. Notice 
the hysteresis  obtained between the curve for
increasing $j$ and the one for decreasing $j$.
Both the junction voltage and ohmic mode voltages are fixed by the length of 
the system. On the other hand the current for which one
obtains these voltages depends on the damping $\alpha$
and strength of the junction $d_a$. For fixed $d_a$, an decrease of
$\alpha$ will tilt the IV curves towards the horizontal voltage axis and 
this will separate the cavity modes and cause large voltage jumps from
a cavity mode to a far away ohmic mode and 
hysteresis. Conversely an increase of $\alpha$ will make the
IV curves more vertical so that the voltage jumps and hysteresis
will be reduced. 
Note also that as in the static case, a magnetic field has no
effect on the IV curves for a single junction.

When the junction is centered as in the left
panel of Fig. \ref{f5} the ohmic modes (resp. junction modes) 
correspond to the 
odd $V=(2 k + 1) \pi/l$ (resp. even $V=2 k \pi/l$) cavity modes so the 
voltage
interval between a junction mode and its corresponding ohmic mode
neighbors is constant. This does not happen when the junction is
off-centered as in the right panel of Fig. \ref{f5} so that one
can get very sharp resonances connecting an ohmic mode to its
junction mode neighbor like the ones for $n=2,3$ and 4. The resonance
for $n=3$ was obtained by decreasing current.
For larger values of $n$ the voltage separation becomes larger so that
the resonances are softer.

\subsection{Study of resonances}

\begin{figure}
\centerline{\epsfig{file=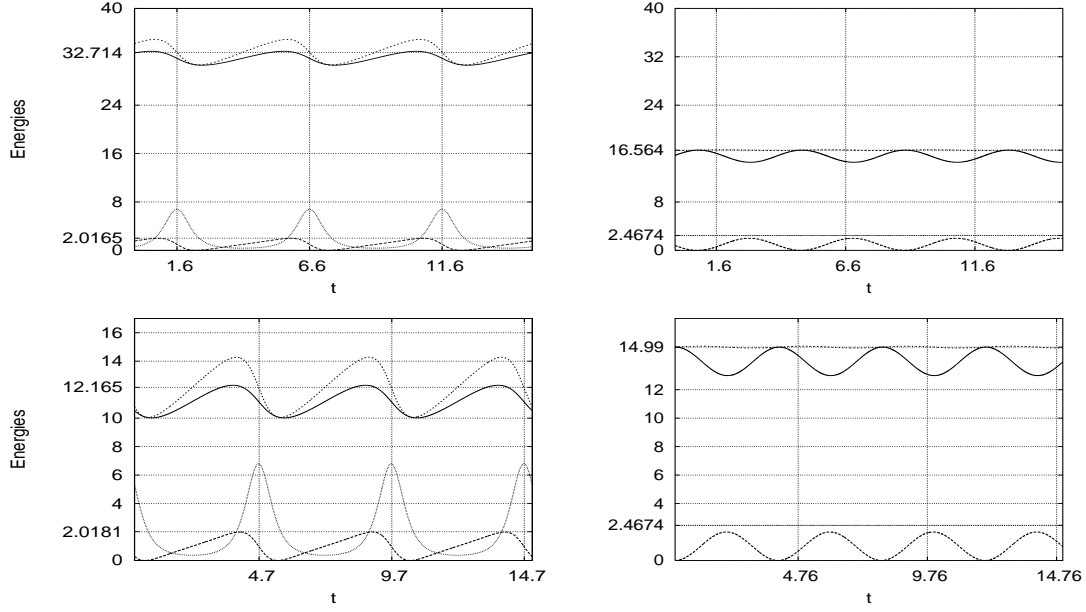,height=16 cm,width=9 cm,angle=270}}
\caption{Plot as a function of time, in each panel, from top to bottom 
of the total energy $E_t$, the energy in the passive region $E_p$, the 
power dissipated by the quasi-particles $P_q$ and the Josephson energy 
$E_j$ for a centered junction $a=5$ (bottom panels) and an off-centered 
junction $a=9$ (top panels). The left panels show the junction mode $n=4$
and the right panels the corresponding ohmic mode for the same current.
The averages of $E_t$ and $E_j$ are shown on the $y$ axis.}
\label{f6}
\end{figure}

We have been able to bound the IV curves for our system. Now let us
discuss the fine structure of the resonances, why some are sharp
(see resonances 3,4 and 6 in Fig. \ref{f5}) while others are not.
To explain this one should consider the energy of the system and
the Fourier modes. Multiplying (\ref{sgd}) by $\phi_t$ and integrating
from $0$ to $l$ we obtain the work equation
\be \label{work}
\partial_t \left[{1 \over 2}\int_{0}^{l} \phi_{t}^2  + \phi_{x}^2 dx 
+ d_a \left(1-\cos\phi(a,t)\right) \right] =
j \int_{0}^{l} \phi_t dx -d_a \alpha\phi_t^2 (a,t).
\ee
We can write (\ref{work}) as
$\partial_t E_t\equiv \partial_t (E_p + E_j) = P_{dc} + P_q $ where
$E_p = {1 \over 2}\int_{0}^{l} ( \phi_{t}^2 + \phi_{x}^2) dx$ 
is the energy in the passive region, $E_j= d_a
\left(1-\cos\phi(a,t)\right) $ the Josephson energy,
$P_{dc} = j \int_{0}^{l} \phi_t dx$ is the power given by the direct 
current 
and $P_q= d_a \alpha\phi_t^2 (a,t)$ is power dissipated by the 
quasiparticles.
In figure \ref{f6}, we plot from top to bottom $E_t,E_p,P_q$ and $E_j$ 
for
a centered junction $a=5$ (bottom panels) and an off-centered junction
$a=9$ (top panels). The left panels show the junction mode $n=4$
and the right panels the corresponding ohmic mode for the same current.
In the junction mode the quasi-particle term $P_q$ oscillates strongly
and the Josephson energy $E_j$ is very anharmonic. In the ohmic mode
$P_q$ is almost constant and $E_j$ is sinusoidal. In all cases the 
energy in the cavity $E_p$ is much larger than the junction energy 
$E_j$.  
When the junction is off-centered (top left panel) the energy stored in
the cavity is much larger. This is due to the smallness of the coupling 
coefficient $c_n$ as we now show.

For that consider the phase $\phi(a,t)$ at the junction as given by 
(\ref{sjm}). It is independent of the junction position $a$. 
Since we are in the junction mode $V=n\pi/l$ we write from 
(\ref{modjonc}) 
$$\phi(a,t) = {n \pi \over l} t + \sum_{p=1}^{+\infty} 
\left [\gamma_p \cos \left(\frac{p n \pi}{l}t\right) + 
\beta_p\sin \left(\frac{p n \pi}{l}t \right) \right ]
c_{np}. $$
The first statement implies that the terms $\gamma_p c_{np}$ 
and $\beta_p c_{np}$ are
constants independent of the junction position $a$ so that 
$\gamma_p$ and $\beta_p$ are proportional to $1/c_{np}$.
As a consequence if $c_{np}$ is small the junction mode 
of index $n$ corresponds to a large amplitude in
the mode $A_{np}$. The system
accumulates this energy in the passive region and the resonance is very
sharp.

Fig. \ref{f6} shows a plot as a function of time, in each panel, 
from top to bottom of the total energy $E_t$, the energy in the
passive region $E_p$, the power dissipated by the quasi-particles $P_q$ 
and
the Josephson energy $E_j$ for
a centered junction $a=5$ (bottom panels) and an off-centered junction
$a=9$ (top panels). 
The left panels show the junction mode $n=4$
and the right panels the corresponding ohmic mode for the same current.
The coupling coefficient is small $c_4=0.3$ in the top panel
while it is maximum $c_4=1$ in the bottom panel where the junction
is centered. Therefore one expects a sharp resonance and a highly
excited passive cavity for $a=9$ as opposed to $a=5$.
This is indeed the case as one can see from the plot of $E_p$ in the
left bottom and left top panels. The Josephson energy $E_j$ and the
power dissipated by the junction are of the same order.
In the ohmic regime this accumulation of energy in the cavity is
absent. There is a small difference in $E_p$ between the two
configurations because they have different lengths of passive regions.
Note that the total energy is practically independent of time and so
is the dissipation. There is an exact balance between the Josephson
energy and the energy in the cavity $E_p$.

\section{The many junction case}

Assuming a general situation with magnetic field, a mixed current feed
and many junctions in the micro-strip we can generalize our model to
\be\label{manyj}
\phi_{tt}- \phi_{xx} + \sum_j d_j \delta(x-a_j) (\sin(\phi) +
\alpha\phi_t)=\nu j \ee
with the boundary conditions $\phi_x|_{x = 0}= b-{jl \over 2}(1-\nu)$
and $\phi_x|_{x = l}= b+{jl \over 2}(1-\nu)$.
If $\nu=1$, 
In the model (\ref{manyj}) $\nu=1$ (resp. $\nu=0$) corresponds to 
an overlap (resp. inline) current feed.

\subsection{Static case}

We have seen that for a single junction the maximum current $j l = d_a$
is reached for any value of the magnetic field. This is not true for
two junctions (a SQUID) where the maximum critical current if obtained 
for
$b=0 ~~{\rm mod.} {2 \pi n \over a_2-a_1}~~$ where $n$ is an integer
\cite{cg04}. Static solutions for many junctions in a 1D microstrip
will be discussed in a forthcoming paper \cite{many}.

\subsection{High voltage case}

As in the case of a single junction, when the stationary
state is reached we can assume $\phi_{t}=V$ and average out the
$\phi_{tt}$ and $\sin(\phi)$ terms yielding the following
boundary value problem for the time independent part $\phi_v$ of $\phi$
\begin{eqnarray} \label{hvolov}
- {\phi_v}_{xx} + \sum_{k=1}^n d_k \delta (x-{a_k})\alpha V = \nu j ,\\
{\phi_v}_{x} |_{0,l}=b \mp {jl \over 2}(1-\nu),
\end{eqnarray}
On each interval separating the junctions ${\phi_v}_{xx} = - \nu j$ so 
that ${\phi_v}$ is a second degree polynomial 
${\phi_v}(x) =-(\nu j/2)x^2 + b x + d$.

Integrating (\ref{hvolov}) from $0$ to $l$ yields the voltage
$$V = \frac{j l}{\alpha\sum_{k=1}^n d_k}.$$
Integrating (\ref{hvolov}) over small domains containing the
junctions we obtain as usual the jump conditions in the derivatives
$$\left[ {\phi_v}_x \right]_{a_i^-}^{a_i^+} = 
\frac{d_i j l}{\sum_{k=1}^n d_k}.$$

Using these remarks we can calculate $\phi_v$ for any number of 
junctions
(for simplicity one can assume $\phi_v(a_1)=0$). Let's introduce 
$S(x)$, the
spatial curvature of $\left<\phi\right>_t$. For two junctions we get
\begin{equation}
\phi_v(x) =
\left\{\begin{array}{l r}
\frac{-\nu j}{2}(x^2 -a_1^2) + \left(b-{jl \over 2}(1 - 
\nu)\right)(x-a_1) 
& x \leq a_1 \\
\frac{-\nu j}{2}(x^2 -a_1^2) + \left({d_1 jl \over d_1+d_2} +
b-{jl \over 2}(1 - \nu)\right)(x-a_1) & a_1<x<a_2 \\
\frac{-\nu j}{2}(x^2 - a_2^2) + \left(b-{jl \over 2}(1+\nu)\right)
(x-a_1) + S_2 & a_2 \leq x \\
\end{array} \right .
\end{equation}
where $S_2 = \frac{-\nu j}{2}(a_2^2 -a_1^2) +
\left({d_1 jl \over d_1+d_2} + b-{jl \over 2}(1 - \nu)\right)(a_2-a_1)$
is the phase difference between the second and first junction 
($S_2 \equiv \phi(a_2,t) -\phi(a_1,t)$). The phase is then given by
\begin{equation}\label{CurveHV}
\phi(x,t) = \phi_v(x) + Vt + C,
\end{equation}{manyj}
where $C$ is an integration constant. Notice that it is possible to 
find $\phi_v(x)$ for an arbitrary number of junctions and compute the
phase difference $\phi_v(a_i)=S_i \equiv \phi(a_i,t)-\phi(a_{i-1},t)$.

\begin{figure}
\centerline{\epsfig{file=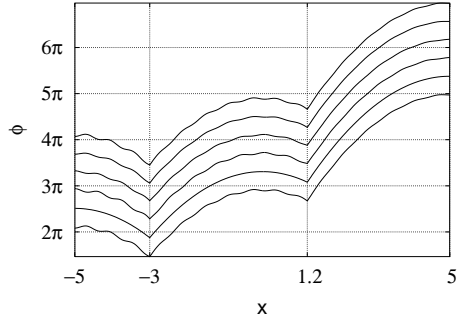,height=6.5 cm,width=4.5 cm,angle=270}}
\caption{High voltage case: plot of $\phi(x,t)$ as a function of
$x$ pour 6 consecutive times separated by $\Delta t=0.1$. The
two junctions are located at $a_1=2$ and $a_2=6.2$. The other 
parameters are $l=10, j=1,$ and $\alpha =0.5$.}
\label{f7}
\end{figure}

In Fig. \ref{f7} we show $\phi(x,t)$ for 6 consecutive values of
time separated by $\delta t=0.1$ for a system of two junctions
located respectively at $a_1=2$ and $a_2=6.2$ in a domain of
length $l=10$ with current $j=1$ fed in overlap geometry ($\nu=1$).
One can see that the behavior
is close to what is predicted by the above expression.

\subsection{Fourier representation} 

We can use the high voltage solution to simplify (\ref{manyj}).
For that take $\psi = \phi + \phi_v$ so that (\ref{manyj}) becomes
\begin{eqnarray}\label{simplj}
\psi_{tt}-\psi_{xx}+\sum_{k=1}^n d_k \delta (x-{a_k})\left (
\sin(\psi-S_k) + \alpha \psi_t-{jl \over \sum_{k=1}^n d_k} \right)=0,\\
\left. \psi_x \right|_{x=0;l} = 0.
\end{eqnarray}
The transformation using the high voltage solution concentrates the
current on the junctions. In this representation inline or overlap 
configurations differ by the terms $S_k$.

Let us go back with Fourier series. Because of the homogeneous Neumann 
boundary conditions on (\ref{simplj}), we can decompose $\psi$ on 
a cosine Fourier series
$$\psi(x,t) = \sum_{n=0}^{+\infty} A_n(t) \cos\left({n\pi x \over 
l}\right)$$ 
and obtain the following equations for the modes $An$
\begin{eqnarray}\label{manya0}
l A_0^{''} + \sum_{j=1}^n d_j \left(\sin(\psi_j - S_j) + \alpha 
\psi_{jt}
-{jl \over \sum_{k=1}^n d_k}\right)=0, \\
A_n^{''} + \left({n \pi \over l}\right)^2 A_n + {2 \over l}
\sum_{j=1}^n d_j c^j_n \left(\sin(\psi_j - S_j) + \alpha \psi_{jt}
-{jl \over \sum_{k=1}^n d_k}\right)=0,
\end{eqnarray}
where $c^j_n = \cos\left({n\pi a_j\over l}\right),~~\psi_j \equiv 
\psi(a_j,t),~~
\psi_{jt} \equiv \partial_t \psi(a_j,t)$.

\section{IV curves for two or more junctions}

\subsection{2 symmetrically placed junctions, limiting behaviors.}

In a symmetric circuit, we have $a_2=l-a_1$ and $d_1=d_2$. For 
simplicity we assume no magnetic field so that
$b=0$ too. With this symmetry, it is easy to show that 
$c_{2k}^2=c_{2k}^1$
and $c_{2k+1}^2=-c_{2k+1}^1$. 
It is important to remark that when $b=0$, $S_2 = 0$. So there is no 
phase shift between the two junctions. Let us now go back 
to the Fourier modes, (\ref{manya0}) becomes:
\begin{equation}\label{A_0_2jsym}
d_1 \left( \sin(\psi_1) + \alpha \psi_{1t}+\sin(\psi_2)+ \alpha
\psi_{2t}\right) -jl = - l A_0^{''}.
\end{equation}
For $p=2k$,
\begin{eqnarray*}\label{A_2k_2j}
A_{2k}^{''} + \left(\frac{2k \pi}{l}\right)^2A_{2k} &=&
-\frac{2 c_{2k}^1}{l}[d_1 (\sin(\psi_1)+ \alpha \psi_{1t}+
\sin(\psi_2)+\alpha \psi_{2t})-jl] \\
&=& 2 c_{2k}^1 A_0^{''}
\end{eqnarray*}
We want to show that the following equation for $A_{2k+1}$ is decoupled 
from the system of Fourier modes
\begin{equation}\label{A_2k+1_2j}
A_{2k+1}^{''}+\left(\frac{(2k+1) \pi}{l}\right)^2A_{2k+1}=
\frac{-2 c_{2k+1}^1}{l}d_1 (\sin(\psi_1) - \sin(\psi_2)
+ \alpha (\psi_{1t} - \psi_{2t})).\end{equation}
For that we group together the even terms and odd terms of the 
Fourier series of $\psi_{1t}$ and $\psi_{2t}$ and define  
$s_o(t)\equiv\sum^{+\infty}_{k=0}
c_{2k}^1 A_{2k}(t)$ and 
$s_e(t) \equiv \sum^{+\infty}_{k=0} c_{2k+1}^1 A_{2k+1}(t)$.
Notice that:
$$\psi(a_1,t)=s_o(t)+s_e(t)~~{\rm and }~~\psi(a_2,t)=s_o(t)-s_e(t)$$
Replacing $\psi_1$ and $\psi_2$ in equation (\ref{A_2k+1_2j}), 
we obtain: 
\begin{equation}\label{decoupl_2j}
A_{2k+1}^{''}+\left(\frac{(2k+1) \pi}{l}\right)^2A_{2k+1}=
\frac{-4 c_{2k+1}^1}{l}d_1 \left(\cos(s_e)\sin(s_o) -
\alpha s_e^{'}\right) 
\end{equation}
This equations show that the system performs like a single centered 
junction with different coupling coefficients. So we can balance 
the Fourier modes in the junction mode: with initial conditions 
$\forall k$, $A_{2k+1}(0)=0$, $A^{'}_{2k+1}(0)=0$ we have 
$S_2(0)=0$ and as a consequence $A_{2k+1}(t) \equiv 0$, 
$\forall t \in \Re^+$. This shows that the equations for the 
odd terms are decoupled from the system.

{\bf Junctions modes.}\\
As we see, the system (\ref{manya0}) behaves like a single 
junction: the odd terms are equal to zero and the even terms 
give the junction modes 
(see (\ref{sjm})):
$$\phi_1=\phi_2=2\arctan\left\{{d_1 \over jl } \left [\tan
\left({1 \over 2 \alpha} V_j \left(t+C_2\right)\right) V_j + 1\right] 
\right\},$$
with $V_j = \sqrt{\left({jl \over d_1}\right)^2-1} = \frac{n \pi}{l}$.

Notice that the two junctions are perfectly synchronized. In a 
non-symmetric situation, junctions modes for both junctions
are particular cases.

\begin{figure} 
\centerline{
\epsfig{file=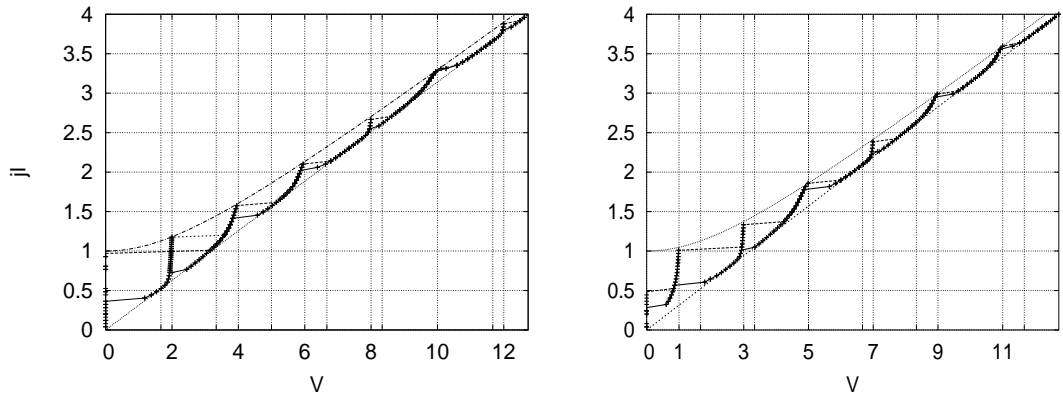,height=16 cm,width= 6 cm,angle=270}}
\caption{IV curves for two symmetrically placed junctions
$a_1=2$ et $a_2=8$.The scale on the $x$ axis is in units of 
$\frac{\pi}{l}$. In the left panel the junctions are identical while
in the right panel the second one is a $\pi$-junction.
Note the resonance on the even harmonics $\frac{2k\pi}{l}$ in the left
panel and on the odd harmonics $\frac{(2k+1)\pi}{l}$ in the right panel.
We could have obtained the same change of IV curves for the
same array of standard junctions by an appropriate choice of the
magnetic field (see text).}
\label{f11}
\end{figure}

{\bf Ohmic modes.}\\
Choosing $S_2 = 0$ allow us to show again analytical solutions. 
Assuming $b=0$, there are three types of solution, given by different 
oscillations of the phase:
\begin{enumerate}
\item The phase oscillates between the junctions so that $V=
\frac{(2k+1)\pi}{a_2-a_1}$. Due to the symmetry of the problem, we 
cannot find solutions with $V=\frac{2k\pi}{a_2-a_1}$.
\item The phase oscillates outside the junctions: 
$V=\frac{(2 p + 1)\pi}{2 a_1}$.
\item The phase oscillates in all the passive parts so that
$V=\frac{(2k+1)\pi}{a_2-a_1}=\frac{(2p+1)\pi}{2 a_1}$. This is
only possible if $a_2-a_1$ is a fraction of $l-a_2$.
\end{enumerate}
To conclude in the symmetric case, we know limit IV curves with exact 
solutions. See the left panel of Fig.\ref{f11}. In the general case, we cannot 
find exact solutions, we have near ohmic mode solutions and peaks at the
expected resonant voltages $V=\frac{n \pi}{l}$.

\subsection{The non-symmetric cases: IV curves }

\begin{figure} 
\centerline{\epsfig{file=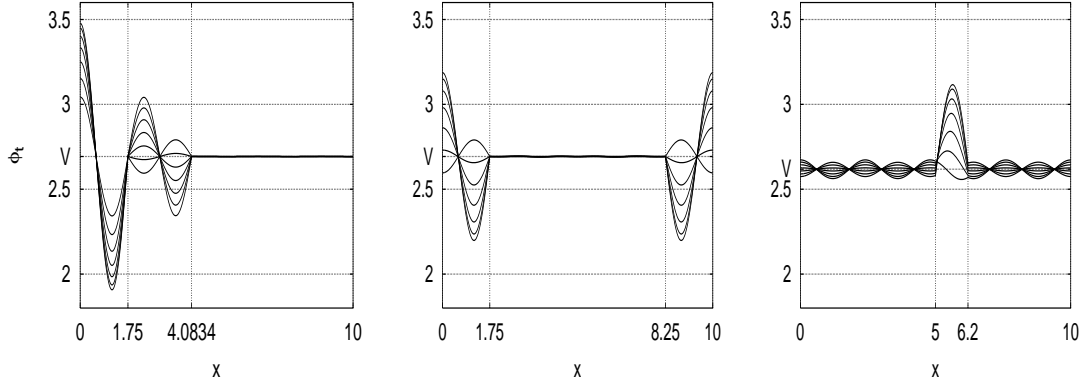,height=16 cm,width=6. cm,angle=270}}
\caption{Three types of ohmic modes obtained for two junctions
in a microstrip. The positions of the junctions are given
in the pictures. We plot $\phi_t\left(x,t\right)$ $\forall x \in [0,l]$ 
for six consecutive times. The parameters of equation 
(\ref{sgd}) are $\lambda=0.5$ and $\alpha=1$. 
All figures have been obtained with the same current $j l=2.6927937$.}
\label{f8}
\end{figure}

The IV curve obtained for a circuit of two junctions placed in 
a non symmetric fashion confirms the existence of ohmic modes 
at the expected voltages. It is still possible to characterize the
IV curves using ohmic modes and junction modes even if there are
only approximate. Without symmetry and when $S_2=0~[\pi]$ we can obtain
exactly any one of ohmic modes discussed in previous section. In 
addition to the symmetric case there two new possibilities where $V$ can 
be computed:
\begin{enumerate}
\item The phase oscillates between the junctions and to the
right of $a_1$. See the left panel of Fig.\ref{f8} as an example.
\item The phase oscillates to the left of $a_2$ and between the
junctions.
\end{enumerate} 
In Fig.\ref{f8}, the middle panel shows the second point of the
symmetric case and the right panel illustrates the case where
the phase oscillates between the junctions for the non-symmetric case.

\begin{figure}
\centerline{\epsfig{file=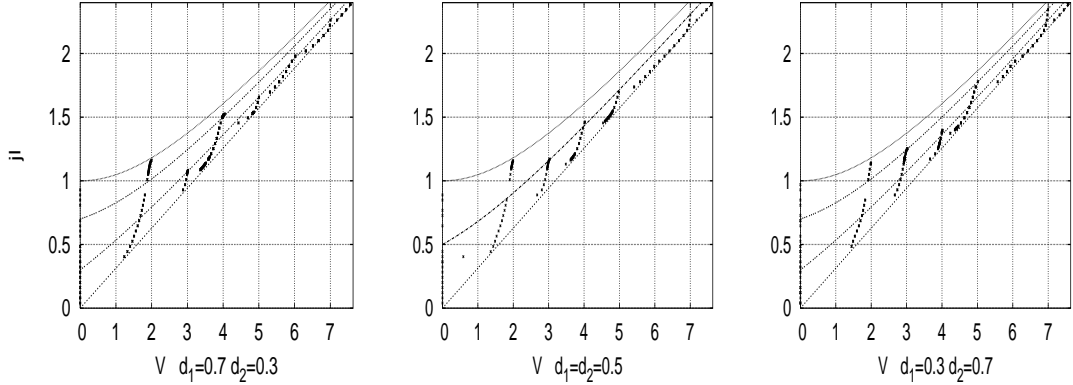, height=16 cm, width=6 cm,angle=270}}
\caption{IV characteristic curves for two junctions placed in $a_1=5$ 
and $a_2=6.2$. The critical current densities are $d_1=0.7,~d_2=0.3$ for 
the left panel, $0.5,~0.5$ for the middle panel and $0.3,~0.7$ for the
right panel. We plot on all figures $j l=\sqrt{V^2+1}$, $j l =V$, 
$j l =d_1\sqrt{V^2+1}+d_2 V$ and $j l =d_2\sqrt{V^2+1}+d_1V$.}
\label{f9}
\end{figure}

Peaks appear at the junction modes $V=\frac{k\pi}{l}$, but they are 
in general smaller than in the symmetric cases. 
Fig.\ref{f9} shows the IV curves for two non symmetrically placed 
junctions and indeed one can see that the height of the resonances
is lower than one for two decoupled junctions (the upper curve in
the three panels). By varying the junction critical current 
(the coefficients $d_1$ and $d_2$ ) we show that the resonances 
are linked to one of the two junctions. To understand 
this phenomenon, go back to the coupling coefficients $c_n^j$.
For the second resonance in Fig. \ref{f9}:
$|c^1_2|=1$, $|c^2_2|=0.73$ so that the junctions are strongly coupled to 
the system. This is like in the symmetric case but it is an exception. The IV 
curve shows that they are together in junction mode (or very close).
Consider now the third resonance. The coupling coefficients are $c^1_3 
= 0$ and $c^2_3 = 0.90$ so that the junction $a_1$ is decoupled from 
the system, we can assume ohmic behavior for it and $a_2$ is in 
junction mode. The fourth resonance corresponds to the reverse 
situation, $c^1_4=1$, $c^2_4=0.06$ where $a_2$ is decoupled.

\begin{figure}
\centerline{\epsfig{file=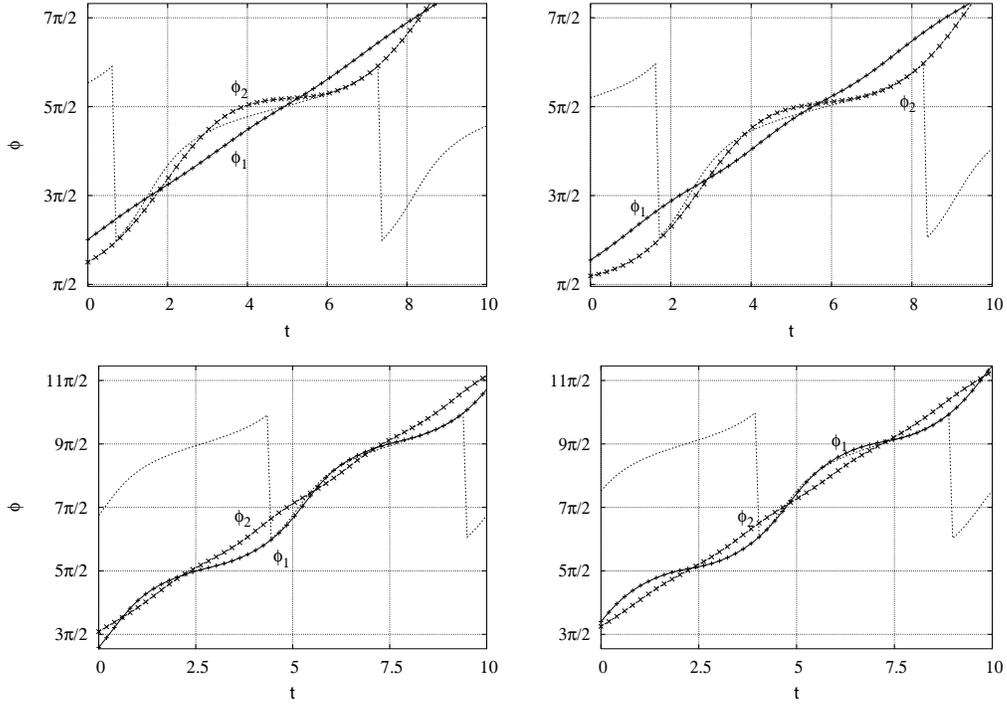,height=15 cm,width=10 cm,angle=270}}
\caption{Plots of the phases at the junctions $\phi(a_1,t)$ and
$\phi(a_2,t)$ as a function of time for the IV curves in the
two left panels of Fig. \ref{f9}. The left panels are
for $d_1=0.7,d_2=0.3$ and the right panels for $d_1=d_2=0.5$.
The top panels show the resonance of order 3 while the bottom
panels show the resonance of order 4. Note how the junctions
switch between ohmic mode and junction mode. The analytic
expressions for the corresponding junction modes (\ref{sjm}) 
are shown in dashed line.}
\label{f10}
\end{figure}

To confirm this, we plot in Fig. \ref{f10}, $\phi_1$ and $\phi_2$ 
for the resonances $3$ and $4$ with $d_1=0.7$, $d_2=0.3$
(left panel of Fig. \ref{f9}) and  $d_1=d_2=0.5$ (middle panel
of Fig. \ref{f9}).
Together with the numerical results we plot the junction mode solution
(\ref{sjm}) in dashed line. For the top panel (third resonance) one
can see that the junction $a_2$ follows closely this behavior as opposed 
to $a_1$ which is in ohmic mode. The situation is reversed for the 
fourth resonance as shown in the bottom panels. 
For resonances 2, 3 and 4, the value of the critical current $d_i~,i=1,2$ does 
not modify appreciably the behavior of the phase at the junctions 
(remark that $S_2 \neq 0$ but stay small).

For a general resonance the junctions will work in different modes,
one will be closer to a junction mode while the other will be closer to 
an ohmic mode. We call this mode a resistor-junction mode.

{\bf Resistor-junction modes.}\\
Resistor-junction modes exist if the circuit has one junction heavily 
coupled while the other is decoupled. In this case, the system behaves almost 
like a single junction circuit. 
This is why these modes are associed to voltages 
$V = \frac{k\pi}{l}$. 

If the junction $a_i$ is in ohmic mode, then $[\phi_x]_{a_i^-}^{a_i^+}=
d_k \sin(Vt+C_1) + C_k$. So that we can only have only one harmonic 
in the circuit. Whereas a junction 
mode exists only with an infinity of harmonics. So the system reaches a 
constructive equilibrium. 
For example in all the panels of Fig. \ref{f10}, we plot the exact junction 
mode to compare. The observation of the phases in the junction mode 
indicates that it lingers more around $(2k+1)\pi/2$ than the 
expression (\ref{sjm}). This gives for the IV curves 
of Fig. \ref{f9} resonances that are slightly higher than expected.
This summarizes the influence of one junction on the other.

\subsection{$\pi$-junctions and magnetic field effects}

We now consider the situation where the critical current $d_i$ for
some $i$ can be negative. Then the ith junction is a so-called $\pi$ junction
for which the Josephson relation is given by $\sin(\phi +\pi)$
instead of $\sin(\phi)$. In this case we will show that the dynamics
is very different than the one where all the $d_i$ are positive.
Therefore combining $\pi$ and standard junctions provides a way 
to change the resonant frequencies of the array.
Another important parameter which we have not varied up to now is the
magnetic field $b$. We will see that a change of $b$ changes the IV 
characteristic in a way that is similar to changing the critical
current density of some of the junctions in the array. This is
another way to tune the array to resonate on specific
frequencies.

We now illustrate these specific points on some examples.
Consider an array of $N$ point junctions described by
$$\phi_{tt}- \phi_{xx} + \sum_{j=1}^{N} d_j \delta(x-a_j) (\sin(\phi) +
\alpha\phi_t + c \phi_{tt})=\nu j~,$$
with the boundary conditions $\phi_x|_{x = 0}= b+{jl \over 2}(1-\nu)$
and $\phi_x|_{x = l}= b-{jl \over 2}(1-\nu)$.
First note that if $d_i<0$ for all i's then we can take 
$\psi= \phi+\pi$ to get us back to the standard case $d_i>0$. We therefore
obtain the same IV curve as when all the $d_i>0$. 

An interesting case corresponds to $N=2$. Let us consider the effect
of an additional magnetic field $b'$ on the IV curve. For that
introduce $\psi(x,t)=\phi(x,t)+f(x)$ where $f(x)= b'(x-a_1)$.
The equation verified by $\psi$ is 
$$\psi_{tt}- \psi_{xx} 
+ d_1 \delta(x-a_1) (\sin(\psi + f_1) + \alpha\psi_t )
+ d_2 \delta(x-a_2) (\sin(\psi + f_2) + \alpha\psi_t )
= \nu j, $$
with the boundary conditions 
$\psi_x|_{x = 0}= b+ b'+{jl \over 2}(1-\nu)$
and $\psi_x|_{x = l}= b+b'-{jl \over 2}(1-\nu)$.
When $f_1=0$ and $f_2=2 \pi$ ie when $b'= 2 \pi / (a_2-a_1)$
we obtain the same equation for 
$\phi$ and $\psi$ apart from the boundary conditions.
The IV curves
for the two magnetic fields $b=0$ and $b= 2 n \pi / (a_2-a_1)$
where $n$ is an integer are then identical. Now if one chooses
a magnetic field $b= \pi / (a_2-a_1)$ then the phases are shifted
by $\pi$ one from the other and this is like changing the sign of
one of the $d_i$. 
In practice
these results only makes sense for magnetic fields for which the
size of the junctions can be neglected. This is the limit of our
approach.

Fig. \ref{f11} shows two IV curves for two symmetrically placed junctions
$a_1=2$ et $a_2=8$ such that
$b=0$ $d_1=d_2=0.5$ corresponding to standard junctions
(left panel) and $b=0$ $d_1=-d_2=0.5$ corresponding to a standard junction
and a $\pi$ junction (right panel). 
We observe in the left panel resonances (junction modes) every
$\frac{2k\pi}{l}$ exactly as for a centered junction. In the right
panel, resonances (junction modes) are found every 
$\frac{(2k+1)\pi}{l}$. The $\pi$ phase shift between
the two junctions implies that odd harmonics are highly coupled and 
even harmonics are decoupled from the system. 
We could have obtained the same
change in the nature of the resonances by using the standard array (left panel)
and changing the magnetic field to $b= \pi/(a_2-a_1)$.  As
the field is changed one then goes continuously from an IV curve 
with only even resonances to an IV curve with only odd resonances.

Now let us consider how these results obtained for $N=2$ affect the
behavior of an array of $N>3$ junctions. If there is a $d$ such that
$a_{i+1}-a_i= n_i d$ where $n_i$ is an integer for all $i$ then
the IV curves for $b=0$ will be the same as for $b= 2\pi/d$.
If one chooses $b= \pi/d$ the junctions such that $f_i=\pi ~{\rm mod.}~ 2 \pi$
will be reversed while the ones for which $f_i=2\pi ~{\rm mod.}~2 \pi$
will not be affected. This is similar to a flute which is tuned
by changing nodes.

\section{Conclusion}

We have introduced a simple and general model using a wave equation 
with delta distributed sine nonlinearities to describe the
dynamics of point Josephson junctions in a 1D microstrip. This
can be extended to $\pi$ junctions for which the current density can
be either positive or negative in the domain. We can also apply it
to situations where the current density in each junction is different.

For a single junction we have shown two limiting behaviors, the ohmic
mode where the junction behaves as a resistor driven by the cavity, separating
waves and the junction mode where the Josephson element is driving the cavity.
These limiting behaviors have allowed us to bound the IV curves and 
understand the observed resonances. When another junction is present 
in the cavity, this simple classification is changed in 
general and we observe ohmic modes and combined junction/ohmic modes. The 
existence or not of the nth resonance is connected to the value of 
the Fourier coefficient $c_n^a = \cos{n \pi a /l}$.

These results carry over to the case of an array with many junctions where
it is possible to choose a voltage where one of the junctions is inactive and 
another is in junction mode. We also use the analysis to understand the
effect of an external magnetic field and the influence of having 0 or $\pi$
junctions. We believe that such a device composed of normal and $\pi$
junctions placed at specific locations in a microstrip 
can be used to great advantage for specific applications like
resonators and frequency mixers.

\section{Appendix}

\subsection{Continuum limit.}

We integrate (\ref{sggh}) from $a-h$ to $a+h$

The first term yields
\begin{eqnarray*}
\int_{a-h}^{a+h} \phi_{tt} dx & = & 
\int_{a-h}^{a+h} \phi_{tt}(a)+(x-a)\phi_{ttx}(a)+O(|x-a|^2) dx\\
& = & 2h\phi_{tt}(a) + 4h^2\phi_{ttx}(a) + O(h^3)\\
\end{eqnarray*}
The second one
$$\int_{a-h}^{a+h} \phi_{xx} dx=\phi_x(a+h)-\phi_x(a-h)$$
and the third one
\begin{eqnarray*}
\int_{a-h}^{a+h}\left [\sin(\phi)+ \alpha\phi_t + c \phi_{tt} \right]
dx & = & \int_{a-h}^{a+h}\sin\left(\phi(a)+(x-a)\phi_x(a)+
O_1(|x-a|^2)\right)\\
& & + \alpha \phi_t(a)+ \alpha (x-a)\phi_{xt}(a) + O_2(|x-a|^2)\\
& & + c\phi_{tt}(a) + c(x-a)\phi_{xtt}(a) + O_3(|x-a|^2)dx\\
& =  & \int_{a-h}^{a+h}\sin\left(\phi(a)+(x-a)\phi_x(a) 
+ O_1(|x-a|^2)\right)dx\\
& & +  \alpha \phi_t(a) + \alpha h \phi_{xt}(a)
 + c\phi_{tt}(a)+ ch \phi_{xtt}(a) + O_4(h^2)\\
\end{eqnarray*}
We now expand the sine term
\begin{eqnarray*}
\sin\left(\phi(a)+(x-a)\phi_x(a)\right)&=&
\sin(\phi(a))\cos((x-a)\phi_x(a))\\
&&+\sin((x-a)\phi_x(a))\cos(\phi(a))
\end{eqnarray*}
so that
\begin{eqnarray*}
&&\int_{a-h}^{a+h}\sin\left(\phi(a)+(x-a)\phi_x(a)\right)\\
&=&\int_{a-h}^{a+h}\sin(\phi(a))\cos((x-a)\phi_x(a))+
\sin((x-a)\phi_x(a))\cos(\phi(a))dx\\
&=&\sin(\phi(a)){2 \over \phi_x(a)} \sin(h\phi_x(a))-0
\end{eqnarray*}
Consider the limit of this term when $h\rightarrow 0$.
We have:
$$ d_a {1 \over 2 h}{2 \over \phi_x(a)}\sin(h\phi_x(a)) \rightarrow d_a,$$
so that the equation for the point junction is
$$ \phi_x(a+h)-\phi_x(a-h) + d_a(\sin(\phi(a))+ \alpha\phi_t(a) +
c \phi_{tt} (a))=j,$$
which is consistent with the model (\ref{sgcd}).
$$\phi_{tt}- \phi_{xx} + d_a\delta(x-a) \left( \sin(\phi) + 
\alpha\phi_t +
c \phi_{tt} \right)=j $$

\subsection{Method of lines}

The basis of the method is to discretize the spatial part of the
operator and keep the temporal part as such. We thereby transform
the partial differential equation into a system of ordinary 
differential equations. This method allows to increase the 
precision of the approximation in time and space independently and
easily. In our case the operator is a distribution so that the
natural way to give it meaning is to integrate it over a volume.
We therefore choose as space discretisation the finite volume
approximation where the operator is integrated over reference volumes.
The value of the function is assumed constant in each volume.

As solver for the system of differential equations, we use
the Runge-Kutta method of order 4-5 introduced by Dormand and Prince
implemented as the Fortran code DOPRI5 by Hairer and Norsett 
\cite{hairer}
which enables to control the local error by varying the time-step.

We first transform (\ref{sgd}) into a system of first order partial
differential equations
We write $\psi(x,t)=\phi_t(x,t)$.
\begin{equation} \label{sysdiscret} 
\left\{  \begin{array}{r c l}
\psi(x,t) & = & \phi_t(x,t) \\
\psi_t(x,t) & = & \phi_{xx}(x,t)-\delta(x-a) (d_a \sin(\phi(x,t)) + 
\alpha \psi(x,t))+j 
\end{array} \right. 
\end{equation}
with the boundary conditions : 
$\phi_x|_{l \over 2}=\phi_x|_{-{l \over 2}}=0$. 

For simplicity we will describe the implementation of 
the finite volume discretisation in the case of a single junction.
We introduce reference
volumes $V_k$ whose centers we call $x_k$, $1\leq k \leq nn$.
The discretisation points are placed such that
the point $x_{ng+1}$ is at the junction, ($x_{ng+1}=a$).
We thus define $x_k$ and $V_k$ using the following identities
$$V_k=\left ]x_k-{h_g \over 2},x_k+{h_g \over 2} 
\right[~,~~~~0<k<ng+1$$
with $(ng+1)h_g=a$
$$V_k=\left ]x_k-{h_d \over 2},x_k+{h_d \over 2} 
\right[~,~~~~ng+1<k<nn+1$$
with $(nn-ng)h_d=l-a$. Finally at the junction, $k=ng+1$
$$V_{k_{ng+1}}=\left ]x_{ng+1}-{h_g \over 2},
x_{ng+1}+{h_d \over 2} \right[.$$
$nn$, $ng$ and $nd$ are respectively the total number of discretisation
points, the number of points to the left of the junction and the number 
of points to the right.

For a fixed t, we assume $\phi(x,t)$ to be constant on each volume $V_k$, 
so that
$$\int_{x_k-{h \over 2}}^{x_k+{h \over 2}} \phi(x,t) dx = 
h \phi(x_k,t)~{\rm,~~with}~h=hg~{\rm or}~h=hd$$
Integrating over $V_k$ yields:
\begin{enumerate}
\item In the linear part of the PDE : $0<k<nn+1$ and $k \neq ng+1$: 
\begin{equation} \label{spacediscr}
\left \{  \begin{array}{r c l}
\psi(x_k,t) & = & \phi_t(x_k,t) \\
\psi_t(x_k,t) & = & 
\frac{\phi(x_{k+1},t)-2\phi(x_k,t)+\phi(x_{k-1},t)}{h^2} + j
\end{array} \right.
\end{equation}
with $h=hg$ for $0<k<ng+1$ or $h=hd$ for $k>ng+1$. We recognize the usual 
discretisation of the second derivative. 

\item At the junction: $k=ng+1$, we obtain
\begin{eqnarray*}\label{disconti}
\int_{x_{ng+1}-{h_g \over 2}}^{x_{ng+1}+{h_d \over 2}}
\delta(x-a)\left(d_a \sin(\phi(x,t)) + \alpha\phi_t(x,t)\right)=\\
d_a \sin(\phi(x_{ng+1},t)) + \alpha\phi_t(x_{ng+1},t)
\end{eqnarray*}
So that the final system is:
\begin{equation} \label{joncdiscr}
\left \{  \begin{array}{r c l}
\psi(x_{ng+1},t) & = & \phi_t(x_{ng+1},t) \nonumber\\
\psi_t(x_{ng+1},t) & = & \frac{4}{hg+hd}\left (\frac{\phi(x_{ng+2},t)-
\phi(x_{ng+1},t)}{hg/2} -\frac{\phi(x_{ng+1},t)-\phi(x_{ng},t)} 
{hd/2} \right)\\
 & & -\frac{2}{hg+hd}\left(d_a \sin(\phi(x_{ng+1},t))
 +\alpha\phi_t(x_{ng+1},t)\right) + j
\end{array} \right.
\end{equation} 
\end{enumerate}
The ODE system (\ref{spacediscr}, \ref{joncdiscr}) is then integrated 
numerically using the DOPRI5 integrator.

\end{document}